\documentclass [%
reprint,
amsmath,amssymb,
aip,
jap,
]{revtex4-1}

\usepackage{graphicx}
\usepackage{amsmath}
\usepackage{amssymb}
\usepackage{bm}
\usepackage{pgf}
\usepackage{color}
\usepackage{amsfonts}
\usepackage{hyperref}
\draft 

\begin{document}


\title{Enhancement of deep-subwavelength band gaps in flat spiral-based phononic metasurfaces using the trampoline phenomena} 


\author{Osama R. Bilal}
\affiliation{Division of Engineering and Applied Science, California Institute of Technology, Pasadena, California 91125, USA}
\affiliation{Department of Mechanical and Process engineering, ETH Zurich, 8092 Zurich, Switzerland}

\author{Andr\'e  Foehr}
\affiliation{Division of Engineering and Applied Science, California Institute of Technology, Pasadena, California 91125, USA}
\affiliation{Department of Mechanical and Process engineering, ETH Zurich, 8092 Zurich, Switzerland}

\author{Chiara Daraio}
\affiliation{Division of Engineering and Applied Science, California Institute of Technology, Pasadena, California 91125, USA}
\date{\today}

\begin{abstract}
	
Elastic and acoustic metamaterials can sculpt dispersion of waves through resonances. In turn, resonances can give rise to negative effective properties, usually localized around the resonance frequencies, which support band gaps at subwavelength frequencies (i.e., below the Bragg-scattering limit). However, the band gaps width correlates strongly with the resonators' mass and volume, which limits their functionality in applications. Trampoline phenomena have been numerically and experimentally shown to broaden the operational frequency ranges of two-dimensional, pillar-based metamaterials through perforation. In this work, we demonstrate trampoline phenomena in lightweight and planar lattices consisting of arrays of Archimedean spirals in unit cells. Spiral-based metamaterials have been shown to support different band gap opening  mechanisms, namely, Bragg-scattering, local resonances and inertia amplification. Here, we numerically analyze and experimentally realize trampoline phenomena in planar metasurfaces for different lattice tessellations. Finally, we carry out a comparative study between trampoline pillars and spirals and show that trampoline spirals outperform the pillars in lightweight, compactness and operational bandwidth.
\end{abstract}

\pacs{}

\maketitle 

\section{Introduction}

Phononic crystals and metamaterials are structured materials that exploit the geometry of their architecture to control the dispersion and the propagation of stress waves. Their operational spectrum can range from a few Hz within the infra-sound range to audible and ultrasonic frequencies\cite{maldovan2013sound}. Phononic crystals and metamaterials have been proposed for different applications, e.g., in  seismic waves' shielding at very low frequencies, and as effective \cite{kim2012seismic, brule2014experiments} noise and vibrations protecting layers in  various frequency ranges \cite{liu2000locally,yang2010acoustic,mei2012dark,xiao2012sound,ma2015purely}. They also have been proposed for frequency filtering \cite{pennec2004tunable,rupp2010switchable}, wave-guiding \cite{Torres_1999,rupp2007design}, computing \cite{sklan2015splash,bilal2017bistable}, subwavelength lensing \cite{ambati2007surface} and acoustic cloaking \cite{torrent2008acoustic}. 

Most phononic crystals and metamaterials consist of basic building blocks that repeat spatially in a periodic or quasi-periodic fashion. One of the important traits of these structured materials is the emergence of band gaps within their frequency dispersion diagrams. Band gaps are frequency ranges where waves are not allowed to propagate within the host medium. The main mechanisms for opening such frequency gaps are Bragg-scattering, local resonance or amplification of inertia. The building blocks are usually composed of one or more materials depending on the desired band gap opening mechanism. To induce a Bragg scattering band gap, the spatial periodicity is usually engineered to match the wavelength of the targeted waves, triggering destructive interferences between traveling and reflected waves \cite{sigalas1993band,kushwaha1993acoustic}. This is usually achieved by having two or more materials within the unit cell or a single material with holes of various shapes. A different path to open band gaps is the presence of locally resonant elements within the building blocks. Such design principle decouples the unit cell size from the wavelength of the attenuated waves and enables subwavelength wave control (i.e., below what is possible through Bragg-scattering)\cite{liu2000locally}. Such resonance-based design principle does not mandate the periodicity of the medium \cite{rupin2014experimental}. Another resonance-based mechanism is the effective amplification of inertia, where a resonator is usually connected to the unit cell through hinges or complaint mechanisms \cite{yilmaz2007phononic}. These resonance-based approaches enable metamaterials to retain properties that do not exist in conventional materials, like negative effective density or stiffness \cite{christensen2015vibrant,cummer2016controlling,ma2016acoustic}. Recently, we presented a platform for realizing different phononic metamaterial physics based on Archimedean spirals spanning Bragg-scattering, local resonance and amplification of inertia utilizing simple variations of the spirals' geometrical parameters and symmetries \cite{foehr2018spiral}.

The ability to control the propagation of elastic waves through the utilization of metasurfaces (i.e., two-dimensional plates) is important for wave guiding or vibration insulation of sensitive equipment \cite{deymier2013acoustic, khelif2015phononic} and the potential realization of meta-devices \cite{zhao2016focusing,bilal2017bistable,cha2018electrical}. Metasurfaces decorated with arrays of pillars \cite{pennec2008low,wu2008evidence} have been utilized in many studies due to their simple geometry \cite{khelif2010locally,rupin2014experimental,assouar2014hybrid,oudich2014negative,pourabolghasem2014experimental,li2015enlargement,jin2016tunable,jin2016phononic,li2016expansion,shu2016band,qureshi2016numerical,guo2017guiding,li2017plate} with applicability across multiple scales \cite{zhao2016focusing, colombi2016forests}. However, similar to most locally-resonant metamaterials, the resonance frequency  of pillared-metasurfaces correlate strongly with the mass and volume of the pillar resonators. Lightweight and planar metasurfaces are useful in various domains, particularly those restricted by mass and volume (e.g., aerospace applications). An additional limitation of resonant metasurfaces is their relatively narrow frequency region of operation. To overcome this obstacle, many approaches are introduced, such as using a multi-material and/or multi-pillars system on the same side of the base plate \cite{zhao2016vibration,zhang2013low}, adding pillars to the bottom and top surface of the base plate \cite{badreddine2012enlargement, zhao2015flexural}, introducing soft material to couple the pillar to the base plate \cite{lixia2016control}, or by introducing holes into the base plate (aka the trampoline phenomenon \cite{bilal2013trampoline}). Trampoline phenomena have been shown to numerically \cite{bilal2013trampoline}  and experimentally \cite{bilal2017observation} increase the band gap width in single material metasurfaces due to the added compliance to the base plate, which  enhances the pillars' resonance. In addition, in a trampoline metasurface, increasing pillar's mass, with the presence of holes, increases the band gap width \cite{coffy2015ultra}. An alternative approach to opening wide band gaps is decorating the base plates with arrays of Archimedean spirals instead of pillars \cite{spadoni2009phononic, bigoni2013elastic, zhu2014negative, foehr2018spiral}.  Metasurfaces realized by patterning arrays of spirals can encapsulate element-wise, real-time tunability \cite{bilal2017reprogrammable} and can be easily produced by additive \cite{bilal2017reprogrammable} or subtractive \cite{bilal2017bistable} manufacturing. Moreover, the tunability of spiral-based metasurfaces \cite{bilal2017bistable,jiang2017dual,bilal2017reprogrammable} has been used to realize all-phononic logic devices \cite{bilal2017bistable}. Finally, the planar nature of the geometry is suitable for miniaturization, for example, by fabricating membranes etched with conventional lithographic techniques \cite{kan2013spiral,liu2018nano}.

 \begin{figure}[b]
 	\begin{center}
 		\includegraphics[scale = 1]{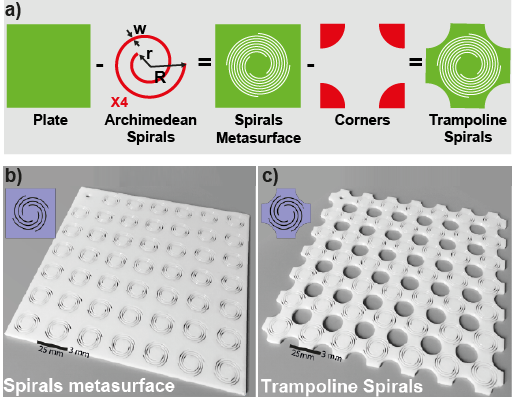}
 	\end{center}
 	\caption{Metasurfaces realization: (a) The construction of spirals-only unit cell by subtracting four concentric Archimedean spirals from a homogeneous plate. The construction of the trampoline spiral unit cell by removing a quarter circle from each corner of the unit cell. (b) Metasurface with concentric Archimedean spirals, consisting of an array of 7 $\times$ 7 unit cells patterned on a polycarbonate plate. The plate thickness is 3.1 mm and the lattice spacing is 25 mm. The spiral inner radius is 4.9 mm and the spiral width is 0.48 mm. (c) Trampoline metasurface composed of spirals and holes with a radius of 7.8 mm. The insets show the unit cells of each metasurface. The scale bar is 25 mm.}
 	\label{fig:Samples}
 \end{figure}
 
 \begin{figure}
 	\begin{center}
 		\includegraphics[scale =1]{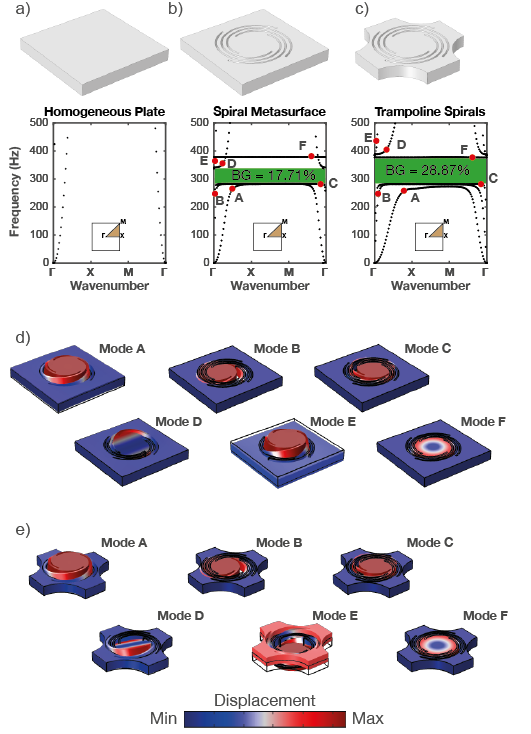}
 	\end{center}
 	\caption{The dispersion curves of three different unit cells in a square lattice: (a) a homogeneous plate, (b) spirals metasurface and (c) trampoline spirals metasurface. The insets represent the symmetry lines for the considered wave vectors along the path $\Gamma-X-M-\Gamma$. The band gap region  are shaded in green. Selected mode shapes around the band gap for (d) spirals metasurface and (e) trampoline spirals. }
 	\label{fig:unitcells_a}
 \end{figure}

 \begin{figure}
 	\begin{center}
 		\includegraphics[scale =1]{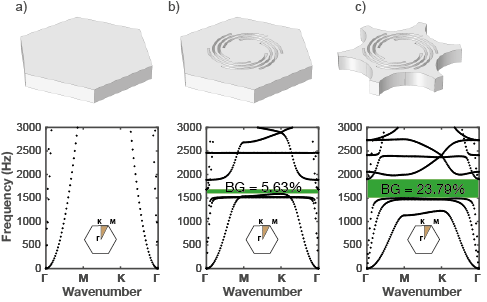}
 	\end{center}
 	\caption{The dispersion curves of three different unit cells in a hexagonal lattice: (a) a homogeneous plate, (b) spirals metasurface and (c) trampoline spirals metasurface. The insets represent the symmetry lines for the considered wave vectors along the path $\Gamma-M-K-\Gamma$. The band gap region  are shaded in green.}
 	\label{fig:unitcells_b}
 \end{figure}
 
In this study, we investigate the effect of the introduction of holes on planar metasurfaces decorated with arrays of spirals. This geometry allows the system to be completely two-dimensional and to reduce the overall mass and volume of the metasurface (Fig. \ref{fig:Samples}). We start our analysis by calculating the numerical dispersion curves, relating frequency to wavenumber, for different arrangements of holes and spirals. Both square and hexagonal packings are considered. We investigate the effect of different spiraling cuts on the width and the position of the band gaps within the frequency spectrum. Then, we analyze the resulting partial and full band gaps, by taking a closer look at their corresponding dispersion curves, for both square and hexagonal lattices. We fabricate two different samples: a spirals-only metasurface and a trampoline-spirals metasurface  (Fig. \ref{fig:Samples} b and c). We experimentally measure the elastic wave propagation characteristics in both samples through different excitations. We consider both in-plane and out-of-plane elastic wave polarizations. Finally, we compare the performance of planar spiraling metasurfaces to pillar-based metasurfaces fabricated with the same material and same base plate dimensions. We investigate the influence of trampoline phenomena on the width of the band gap in both configurations.

\section{Numerical simulations}
We consider an infinite array of repeating unit cells in both $x$ and $y$ directions. The basic building block is a single material plate with side length \textit{a} and thickness \textit{th} carved with four concentric Archimedean spirals. The elastic wave equations for a heterogeneous medium is \cite{Bilal_PRE_2011}:	
\begin{equation}
	\nabla \ldotp \mathbf{C} \colon \frac{1}{2}(\nabla \mathbf{u} + (\mathbf{u})^{\textrm{T}}) = \rho \mathbf{\ddot{u},}
\end{equation}	
where $\nabla$ is the gradient operator, $\mathbf{C}$ is the elasticity tensor, $\mathbf{u}$ is the displacement vector, $\rho$ is the density, and $(.)^{\textrm{T}}$ is the transpose operation. To obtain the dispersion diagram correlating frequency and wave number for our material, we apply the Bloch wave formulation in both $x$ and $y$ directions  (i.e., Bloch boundary conditions) \cite{brillouin1953wave}. The Bloch solution is assumed to be $\mathbf{u(x,\kappa;t) = \tilde{u}(x,\kappa)}e^{i(\mathbf{\kappa.x}-\omega \mathbf{t})}$ where $\tilde{\mathbf{u}}$ is the Bloch displacement vector, $\mathbf{\kappa}$ is the wave vector, $\omega$ is the frequency, $\mathbf{x} = \{x, y, z\}$ is the position vector, and $t$ is time. This form of solution yields a complex eigenvalue problem when plugged into the wave equation in a discretized form:
\begin{equation}
	\lbrack\mathbf{K}(\kappa) - \omega^2 \mathbf{M}\rbrack\mathbf{u} = 0
\end{equation} 
where $\mathbf{K,M}$ are the stiffness and mass matrices.  We solve the complex eigenvalue problem using the finite element method.	

\begin{figure}
	\begin{center}
		\includegraphics[scale =1]{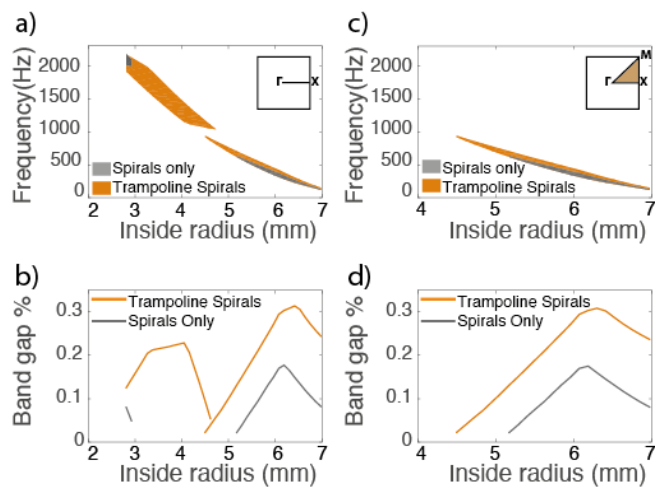}
	\end{center}
	\caption{First band gap evolution for a square lattice: The first band gap edges as a function of inner radius for both spirals (gray) and trampoline spirals (orange) metasurfaces along (a) $\Gamma-X$ direction, partial band gaps (c) $\Gamma-X-M-\Gamma$, full band gaps. The percentages of the first band gaps for (b) partial band gaps and (d) full band gaps. The insets in (a) and (b) represent the symmetry line(s) for the considered wave vectors.}
	\label{fig:sqr_agg}
\end{figure}

We numerically analyze two configurations of unit cells. The first unit cell configuration is constructed by cutting four concentric Archimedean spirals from a homogeneous plate. The second unit cell is constructed from the same spiraling pattern with additional circular cuts at the corners of the unit cell (Fig. \ref{fig:Samples} a). We refer to the first configuration as spirals-only metasurface and to the second as spiral-trampoline metasurface. The polar representation of the Archimedean spiral is: $r(s)= R-(R-r)\,s,~ \phi (s) = 2\pi\, n\, s,$ where $r$ is the inside radius, $R$ is the outside radius, $n$ is the number of turns and $s \in [0; 1]$. The repetition of Archimedean spirals can give rise to a plethora of intriguing wave phenomena depending on the underlying lattice vectors (for example, frequency dependent wave beaming) \cite{foehr2018spiral}. In our study, we consider both square and hexagonal lattice tessellations. We start our analysis by comparing the dispersion diagrams for both metasurfaces configurations to a homogeneous unit cell with the same dimensions as a reference (Figs. \ref{fig:unitcells_a} and \ref{fig:unitcells_b}). We vary the wavenumber ($\kappa$) along the symmetry lines  $\Gamma-X-M-\Gamma$ for the square lattice case (Fig. \ref{fig:unitcells_a}) and along the vectors $\Gamma-M-K-\Gamma$ for hexagonal lattices (Fig. \ref{fig:unitcells_b}). The lattice constant, defined as the distance between the center of two neighboring unit cells, is $a = 25~ mm$. The parameters for the spiral geometry in both lattices are: lattice constant, $a = 25$ mm, thickness, $th = 3.1$ mm, spiral width $w = .48$ mm, hole radius and spiral outside radius, $r = 8.1~ mm$, spiral inside radius, $r_{in} = 5.9~ mm$. The material parameters are \cite{bilal2017observation} ($\rho = 1200~ Kg/m^3$, $E = 2.3~ GPa$, $\nu = 0.35$). The resulting dispersion curves are plotted in figures \ref{fig:unitcells_a} and \ref{fig:unitcells_b}.
 
\begin{figure*}
	\begin{center}
		\includegraphics[scale =1]{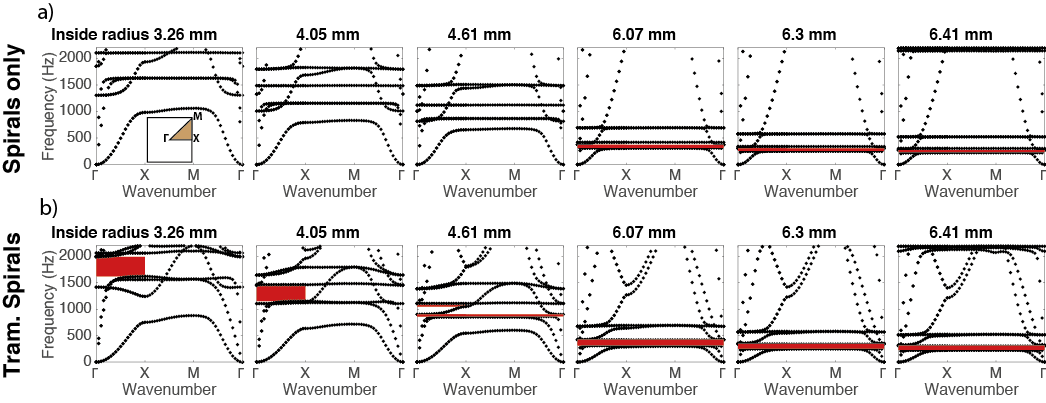}
	\end{center}
	\caption{Band gap evolution for square packing of spirals as a function of its inner radius for (a) spirals-only configuration and (b) trampoline spirals. The inset represents the symmetry lines for the considered wave vectors. Band gap frequency ranges are highlighted in red. }
	\label{fig:sqr_det}
\end{figure*}

In the square lattice case, the introduction of the spiral pattern opens a  band gap (Fig. \ref{fig:unitcells_a} b) with normalized width $\Delta \omega / \omega_{c}$ = 17.71\% (where $\omega_{c}$ is the band gap central frequency). Using this percentage metric takes into account both the absolute width and the central frequency of the gap. After perforation (i.e., introduction of the holes), the same spiraling geometry retains a 28.87\% normalized band gap width (Fig. \ref{fig:unitcells_a} c) with an increase of 63\% from the spirals-only metasurface. To highlight the trampoline effect, we consider the vibrational mode shapes of the unit cell and compare the spiral-metasurface modes to the trampoline-spiral modes. The first three fundamental vibrational modes, namely the out-of-plane mode of the unit cell and the two in-plane modes, are plotted in the first row in panels d for the spiral metasurface and in panel e for the trampoline spiral metasurface. The introduction of holes in the trampoline metasurface does not change the frequency or the shape of either of the first three modes (modes A-C) and therefore does not change the lower edge of the band gap for the given set of spiral parameters. The upper edge of the gap, however, is shifted upwards due to the presence of the holes. The change of the position of the mode shapes in the frequency spectrum can be observed in both modes (D) and (E) in figure 2(panel d and e). In particular, the (E) mode in the trampoline case shows the engagement of the base plate in the resonance motion of the spiral core of the unit cell, which highlights the signature of the trampoline effect \cite{bilal2013trampoline}. Moreover, it is worth noting that mode (F), which is a rotational mode of the spiral core does not change position with or without the presence of the holes. This fixed position of the rotational mode (F) makes it is easy to note the change in the frequency of modes (D) and (E) as they switch from being at a lower frequency to a higher frequency relative to mode (F).

In the case of the hexagonal lattice, the same spiral pattern opens a narrow band gap (Fig. \ref{fig:unitcells_b} b) with a normalized width of 5.63\%. The introduction of holes increases the band gap relative width to 23.79\% with an increase of 322\%.  All the band gaps reported in figures  \ref{fig:unitcells_a} and \ref{fig:unitcells_b}  are in the deep-subwavelength frequency range(i.e., below the Bragg scattering limit)\cite{foehr2018spiral}, in comparison to the homogeneous plate properties. For the considered unit cells, the square lattice band gaps are lower in frequency than the hexagonal ones by approximately a factor of 5. 

\begin{figure}[b]
	\begin{center}
		\includegraphics[scale =1]{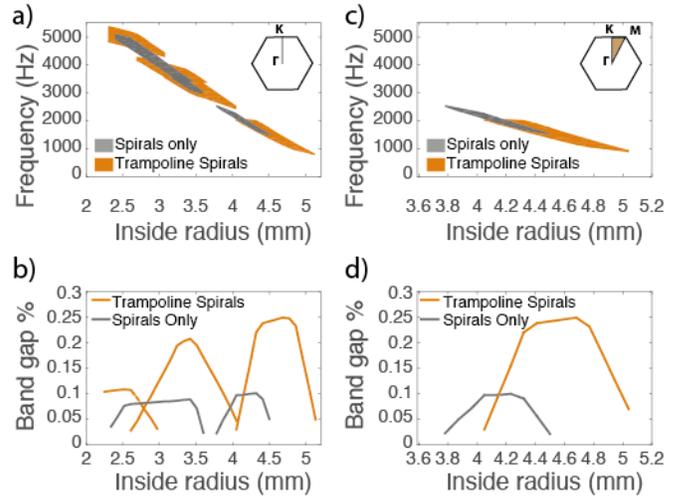}
	\end{center}
	\caption{First band gap evolution for a hexagonal lattice: The first band gap edges as a function of inner radius for both spirals (gray) and trampoline spirals (orange) metasurfaces along (a) $\Gamma-X$ direction, partial band gaps (c) $\Gamma-X-M-\Gamma$, full band gaps. The percentages of the first band gaps for (b) partial band gaps and (d) full band gaps.}
	\label{fig:hex_agg}
\end{figure}

\begin{figure*}
	\begin{center}
		\includegraphics[scale =1]{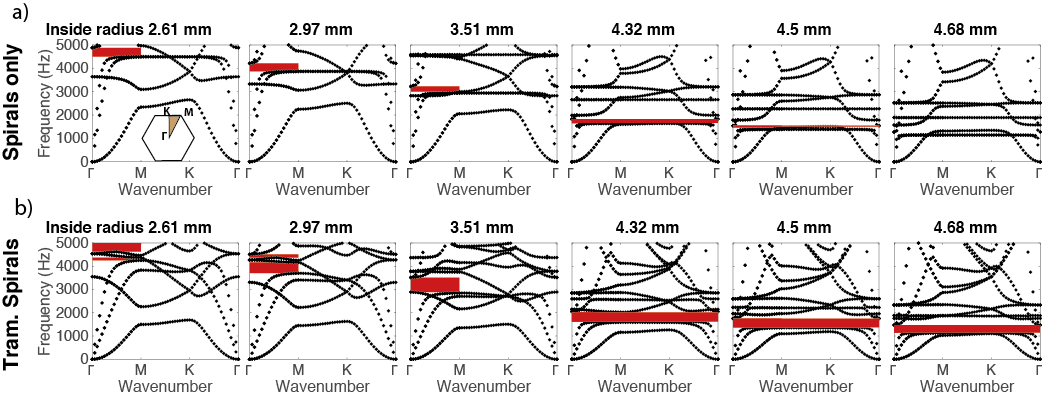}
	\end{center}
	\caption{ Band gap evolution for hexagonal packing of spirals as a function of its inner radius for (a) spirals-only configuration and (b) trampoline spirals. The inset represents the symmetry lines for the considered wave vectors. Band gaps frequency ranges are highlighted in red.}
	\label{fig:hex_det}
\end{figure*}

To analyze the influence of the trampoline phenomena on the relative band gap width of spiraling metasurfaces, we systematically vary the inner radius of the spirals ($r_{in}$) from 2.8 mm to 7 mm (Fig. \ref{fig:sqr_agg}). We record the frequencies of the upper and lower edges of the first band gaps for both spirals-only and spiral-trampoline metasurfaces (Fig. \ref{fig:sqr_agg} a,b). We first consider the partial band gaps, i.e., focusing only on the waves propagating along the $\Gamma-X$ direction, for both configurations (Fig. \ref{fig:sqr_agg} a). The evolution of the first partial band gap as a function of $r_{in}$ is divided in two regions. The first region includes $r_{in}$ ranging from 2.8 mm to 4.7 mm. Spirals-only metasurfaces have no significant band gaps, while trampoline-spiral metasurfaces have a maximum band gap relative width of 22.85\%. To understand the reasoning behind the emergence of the band gaps in the trampoline configuration, we plot the dispersion curves of selected $r_{in}$ values (Fig. \ref{fig:sqr_det}). An increase in the compliance  ``softening" of the plate base affects the lower limit of the Bragg-scattering frequency\cite{foehr2018spiral}, which gives rise to the partial band gaps. Such a phenomenon can be observed in figure \ref{fig:sqr_det} a vs. b at $r_{in}$ = 3.26. As the inner radius of the spirals increases (e.g., $r_{in}$ = 4.61), the central mass of the spiral gets heavier. The increase of the relative mass of the core of the spiral lowers the resonance frequency without any significant effect on the overall stiffness of the plate, which narrows width of the partial band gap while giving rise to a full one (Fig. \ref{fig:sqr_det} b).

The region of the second partial band gap extends between $r_{in}$ = 4.8 mm until the end of the considered parametric sweep at $r_{in}$ = 7 mm. In this region both partial and full band gaps coincide, as the resonance induced by the spiral core is  strong enough  to open a full band gap starting from 118 Hz. A full band gap starts at $r_{in}$ = 4.5 mm for the trampoline spirals, but not until $r_{in}$ = 5.1 mm for the spirals-only configuration (Fig. \ref{fig:sqr_agg} c). The maximum full band gap in the trampoline case exists at $r_{in}$ = 6.3 mm with normalized width of 30.85\%. That is almost as twice as the maximum gap for spirals-only at $r_{in}$ = 6.2 mm which peaks at 17.72\%. The lowest gap for the considered parameters spans the range of 118 - 150 Hz for trampoline spirals, with a three-fold increase over spirals-only metasurfaces (Fig. \ref{fig:sqr_agg} d).  As a conclusion for the square lattice configuration,  the lower edge of the band gap in the spiral-trampoline case is always below the spirals-only configuration. In addition, the gaps in the spirals-trampoline case are always wider. Having band gaps starting at lower frequencies translates to smaller unit cell sizes, in comparison to spirals-only metasurfaces, for the same operating frequency. The increased width in band gap is beneficial as it translates to larger operational bandwidth.

It is established that periodicity is not essential for opening locally resonant or inertially amplified band gaps. However, the addition of ordered holes, inducing Bragg-scattering hybridization, can give rise to different phenomena depending on the lattice configuration. To fully capture the influence of lattice configuration on trampoline phenomena, we consider hexagonal packing of spiraling metasurfaces with perforation (i.e., introduction of the holes) at the six corners of the unitcell (Fig. \ref{fig:unitcells_b} c). We record the evolution of the band gap width as a function of the spiral inner radius (Fig. \ref{fig:hex_agg}). We vary $r_{in}$ from 2.2 to 5.2 mm. The band gaps for spirals-only metasurfaces evolve in two separate regions, similar to that of the square lattice configuration. The first region represents partial band gaps at $r_{in}$ = 2.3 to 3.6 mm, while the second region corresponds to full band gaps at $r_{in}$ = 3.78 to 4.5 mm (Fig. \ref{fig:hex_agg} a and c). To examine the emergence of the band gaps in the trampoline configuration, we plot the dispersion curves of selected $r_{in}$ values (Fig. \ref{fig:sqr_det}). The band gap opening mechanism is pure Bragg-scattering from the beginning of the parameter sweep up to $r_{in}$ = 2.61 mm, where a hybrid band gap starts to appear. The partial gap starts as a hybridization between Bragg-scattering and resonance at $r_{in}$ = 2.61 mm (Fig. \ref{fig:hex_det} a) below the Bragg-scattering gap. The hybridization is more pronounced in the trampoline configuration compared to that of the spirals-only metasurface (Fig. \ref{fig:hex_det} b). As the inner radius increases, the mass of the  spiral core increases, which lowers its resonance frequency, causing the lower band gap to overtake the pure Bragg-scattering band gap at $r_{in}$ = 3 mm. The hybrid gap peaks at $r_{in}$ = 3.4 mm, after which it shifts to lower frequencies as the resonance increases. The resonance eventually dominates and opens a locally resonant full band gap starting at $r_{in}$ = 4.05 mm for the trampoline-spiral metasurfaces. The maximum normalized width of the full band gap in the trampoline-spiral metasurfaces is 24.89\% taking place at $r_{in}$ = 4.68 mm, which is 150\% more than the 9.98\% maximum full band gap for spirals-only metasurfaces at $r_{in}$ = 4.23 mm. Moreover, the lowest possible full band gap with spirals only starts at 1511 Hz, which is almost twice that of the lowest  full band gap for trampoline-spirals starting at 878 Hz. As a conclusion for both square and hexagonal lattice configurations, adding perforation to the base plate increases the width of both partial and full band gaps in all polarizations. 

\section{Experiments}

\begin{figure}[h]					
	\begin{center}
		\includegraphics[scale =1]{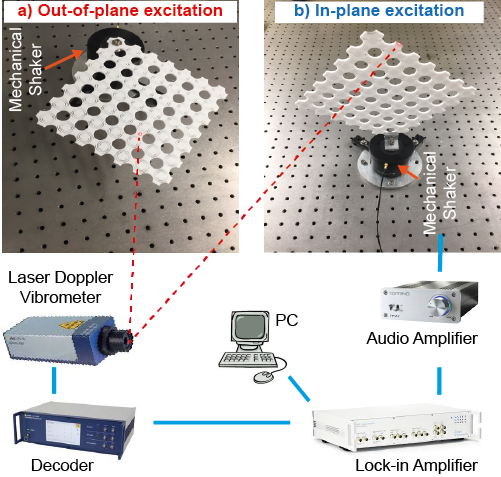}
	\end{center}
	\caption{Experimental setup: (a) A metasurface sample mounted horizontally on a mechanical shaker to test out-of-plane (bending) waves. (b) The same metasurface sample mounted on the shaker vertically to test in-plane waves. In both panels the red dot is the laser Doppler vibrometer measurement point.}
	\label{fig:exper}	
\end{figure}

To experimentally verify the existence of deep subwavelength band gaps within our metasurfaces (both spirals-only and trampoline-spirals), we fabricate an array of 7x7 unit cells made of Polycarbonate (PC) using a Fortus 400mc from Stratasys (Fig. \ref{fig:Samples}). The parameters for the fabricated geometry are: lattice constant, $a = 25$ mm, thickness, $th = 3.1$ mm, spiral width $w = .48$ mm, hole radius and spiral outside radius, $r = 7.8~ mm$, spiral inside radius, $r_{in}$ = 4.9 mm. We excite the metasurfaces with a harmonic signal (i) out-of-plane, perpendicular to the metasurfaces (Fig. \ref{fig:exper} a) and (ii) in-plane, along the thickness of the metasurfaces (Fig. \ref{fig:exper} b). The excitation point in both cases is at one of the plate corners using a mechanical shaker (Bruel \& Kjaer Type 4810). The excitation signal is sent to the shaker from the computer through an audio amplifier (Topping TP22). The traveling wave velocity in the plate is detected by a laser Doppler vibrometer (Polytec OFV- 505 with a OFV-5000 decoder). The velocity is sent back to the computer through a lock-in amplifier from Zurich Instruments (HF2LI). We vary the excitation frequency from 50 Hz to 1.5 kHz in 3 Hz increments and record the wave transmission through the metasurfaces measured at the points illustrated by the red laser dot path in fig 7 in figure \ref{fig:exper}.

\begin{figure*}
	\begin{center}
		\includegraphics[scale =1]{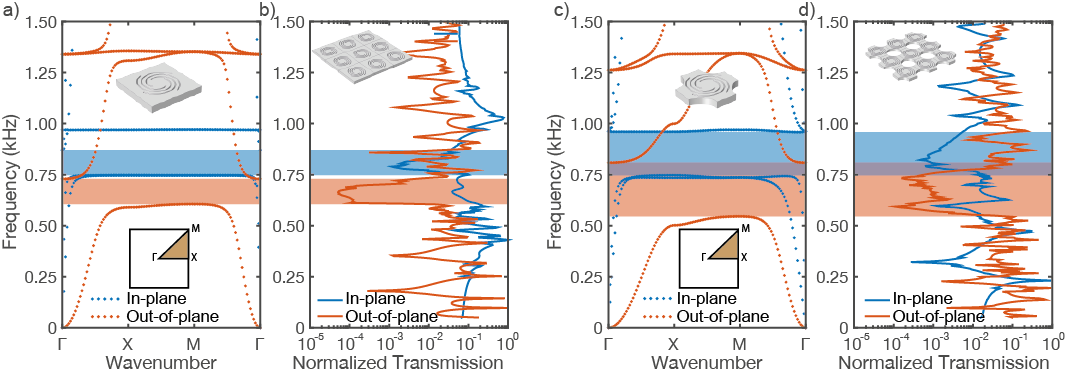}
	\end{center}
	\caption{Experimental and numerical characterization of metasurface: Dispersion curves of (a) spirals-only metasurface and (b) its measured frequency response function. (c) Dispersion curves of spiral-trampoline metasurface with the same plate thickness and spiral parameter. (d) The measured frequency response function. The lines are color-coded based on polarization; blue for in-plane and orange for out-of-plane. The band gaps are highlighted in blue for in-plane and orange for out-of-plane.}
	\label{fig:Exp}
\end{figure*}

The dispersion curves for both spirals-only and trampoline-spiral metasurfaces are plotted in figure \ref{fig:Exp} a and c, respectively. Since each polarization is excited separately, the dispersion lines are colored according to the polarization; blue for in-plane waves and orange for out-of-plane waves. The spirals-only geometry retains a separate band gap for each polarization (Fig. \ref{fig:Exp} a). The in-plane band gap ranges from 604 Hz to 730 Hz, while the out-of-plane band gap ranges from 744 Hz to 870 Hz. The trampoline-spiral geometry has a band gap from 545 Hz to 808 Hz for in-plane waves and another band gap from 744 Hz to 960 Hz for out-of-plane waves (Fig. \ref{fig:Exp} c). The two gaps have a small intersecting frequency range that opens a full band gap for both in-plane and out-of-plane polarizations.

The recorded wave velocities at the measurement points, which are highlighted in red in figure \ref{fig:exper}, are normalized by the measured velocities at the excitation point. To calculate the transmission, the recorded wave velocities at the measurement points (highlighted red dots in fig 7) are normalized by the measured velocities at the excitation point. The measurements are done separately for each wave polarization. The frequency response function correlating the frequency of excitation and the normalized transmission amplitude for both in-plane (blue) and out-of-plane (orange) waves  are plotted in (Fig. \ref{fig:Exp} b and d). In the spirals-only case, we observe a perfect match for out-of-plane waves, while the measured in-plane gap is slightly smaller than predicted (Fig. \ref{fig:Exp} b). In the trampoline case, the out-of-plane gap's upper edge is slightly lower than numerically predicted. In general, the experimentally measured band gaps for out-of plane (bending) waves in both metasurfaces are in good agreement with the numerical prediction, (Fig. \ref{fig:Exp} b), while the in-plane waves have slightly higher frequencies than predicted. This deviation for the in-plane gaps could be due to fabrication imperfections of the cutting width of the spirals, which affect in-plane waves more than out-of-plane waves \cite{foehr2018spiral}. For both polarizations, the addition of holes significantly increased the width of the gap in both numerics and experiments.

\begin{figure*}
	\begin{center}
		\includegraphics[scale =1]{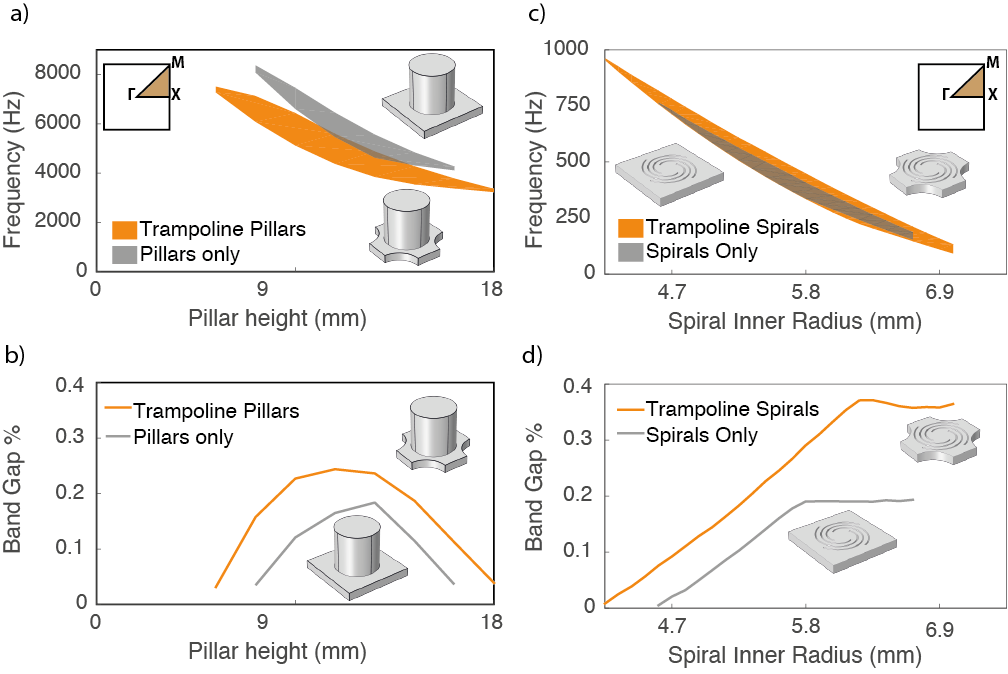}
	\end{center}
	\caption{Numerical comparison between the first complete band gap induced by different pillar and spiral configurations. Pillared metasurface: a) Frequency range of the complete band gap with different pillar heights in pillared metasurface and trampoline pillared metasurface. b) Band gap percentage of both pillared configurations as a function of pillar height. Spiral-based metasurface: c) Frequency range of the complete band gap with different spiral inner radius in trampoline versus non-trampoline metasurface. d) Band gap percentage of both spiral configurations as a function of inner radius.}
	\label{pillars_vs_spirals}
\end{figure*}

\section{Trampoline pillars versus trampoline spirals}

Finally, we analyze the trampoline effect on different metasurface configurations by comparing the  full band gap width resulting from the erection of pillars on a plate (pillar-based metasurfaces) and a planar plate with spiraling cuts (spiral-based metasurfaces). We numerically simulate different variations of both geometries (pillars and spirals) and record the frequency range of the first full band gap for each. We also consider the percentage corresponding to the normalized band gap width (Fig. \ref{pillars_vs_spirals}). All geometries are simulated with the same parameters as in Bilal \textit{et al.,} \cite{bilal2017observation} using ABS plastic ($\rho = 1040~ Kg/m^3$, $E = 1.65~ GPa$, $\nu = 0.35$), square lattice spacing $a = 25~ mm$, and a plate thickness $th = 3.2~ mm$. We choose the outer radius of the spirals, the pillars and the holes to be identical with $r = 7.8~ mm$.

For the considered pillar-based metasurfaces, the addition of holes lowers both the upper and lower edges of the band gap. Trampoline effect slightly expands the existence of full band gaps as a function of pillar heights in both directions (Fig. \ref{pillars_vs_spirals} a). Having band gaps at lower frequencies, even with the same width $\Delta \omega$, results in a higher band gap percentage BG = $\Delta \omega / \omega_{c}$; because the central frequency of the band gap $\omega_{c}$ decreases. The maximum gap percentage for the pillars-only configuration is $18.4\%$, while the band gap percentage after adding the holes can go up to $24.4\%$ with an increase of $32.6\%$ (Fig. \ref{pillars_vs_spirals} b). In the spirals configuration, the addition of holes increases the frequency of the upper band gap edge, however, with limited influence on the lower edge of the gap. The trampoline effect also expands the inner radii range of metasurfaces with full band gaps in both directions (Fig. \ref{pillars_vs_spirals} c). The maximum gap percentage for the spirals-only configuration is $19.4\%$, while the band gap percentage after adding the holes can go up to $37.1\%$ with an increase of $91.3\%$ (Fig. \ref{pillars_vs_spirals} d). 

For both spiraling and pillared metasurfaces, perforation increases the percentage of the band gaps. However, the effect in the spirals case is more profound with almost double of the maximum possible band gap relative width. It is worth noting that the lowest frequency for the bottom edge of the band gap is 3230 Hz in the trampoline-pillars case, while being 125 Hz for trampoline-spirals with the same spacing. That translates to a factor of 25 in operational frequency in addition to more than an order of magnitude increase in band gap percentage (3\% BG at pillar height 18 mm and 36.8\% BG at inner radius 7 mm). In addition, spiraling metasurfaces (with or without the holes) retain both mass and volume advantages over pillared metasurfaces.

\section{Conclusion}
In this paper, we introduced the concept of linear, local-resonance enhancement (trampoline effect) to planer metasurfaces carved with Archimedean spirals. We first numerically analyze the effect of the increase of the inner radius of the spiral on the band gap frequency range and the normalized band gap percentage. We consider both partial (directional) and complete band gaps. Then, we explore the effect of the underlying lattice on trampoline metasurfaces by analyzing hexagonal packing of spirals with and without holes. In a square lattice, the trampoline effect for spiral based metasurfaces leads to the opening of full and partial band gaps where spiral-based metasurfaces (made out of the same material and spiral parameters) do not support band gaps. In a hexagonal lattice, there exists a small range of parameters where a spirals-only metasurface can open band gaps while trampoline metasurfaces can not. Generally, the spiral trampoline metasurface outperforms the spirals-only metasurface in band gap percentage. In order to validate the numerical analysis, we fabricate two metasurfaces -spirals only and trampoline spirals- using a single material through additive manufacturing. Both samples are exited harmonically at the corner using a mechanical shaker in both in-plane and out-of-plane polarization. The experimentally observed band gap frequency ranges agree well with our numerical predictions.  Moreover, we compare the band gap width of pillared metasurfaces against spiral-based metasurfaces, in both the absence and the presence of the trampoline effect, all fabricated from same material. In the case of the trampoline spirals, the band gaps are wider with significantly less mass. Such properties could be beneficial in aerospace vibration insulation and naval domains, where limitations on the overall system mass and volume are present.  
    
\begin{acknowledgments}
We are grateful for T.~Jung's help with the additive manufacturing. O. R. Bilal acknowledges the support from the ETH Postdoctoral Fellowship FEL-26 15-2. This work was partially supported by ETH grant No. ETH-­24 15-2. 
\end{acknowledgments}

\bibliography{Trampoline_Spirals_v3}

\end{document}